\documentclass[11pt,a4paper]{article}
\usepackage{graphicx}
\usepackage{multirow}
\usepackage{url}
\usepackage{amssymb}

\newcommand{\A}{\mathbf{A}}
\newcommand{\m}{\mathbf{m}}
\newcommand{\data}{\mathbf{d}}
\newcommand{\w}{\mathbf{w}}
\newcommand{\W}{\mathbf{W}}
\newcommand{\n}{\mathbf{n}}
\newcommand{\vv}{\mathbf{v}}
\newcommand{\rr}{\mathbf{r}}
\newcommand{\uu}{\mathbf{u}}
\newcommand{\D}{\mathbf{D}}
\newcommand{\zero}{\mathbf{0}}

\title{Nonlinear regularization techniques for seismic tomography}
\author{I. Loris, H. Douma,G. Nolet, I. Daubechies}

\author{I. Loris${}^1$, H. Douma${}^2$, G. Nolet$^{3}$, I. Daubechies${}^4$, C. Regone${}^5$\\
${}^1$Mathematics Department,Vrije Universiteit Brussel\\
${}^2$Department of Geosciences, Princeton University\\
${}^3$Geosciences Azur, Universit\'e de Nice-Sophia Antipolis,CNRS/IRD\\
${}^4$Program in Applied and Computational Mathematics, Princeton University\\
${}^5$BP America Inc., 501 Westlake Park Blvd, Houston, TX 77079}
\date{}

\begin{document}

\maketitle

\begin{abstract}
The effects of several nonlinear regularization techniques are
discussed in the framework of 3D seismic tomography.
Traditional, linear, $\ell_2$ penalties are compared to
so-called sparsity promoting $\ell_1$ and $\ell_0$ penalties,
and a total variation penalty. Which of these algorithms is
judged optimal depends on the specific requirements of the
scientific experiment. If the correct reproduction of model
amplitudes is important, classical damping towards a smooth
model using an $\ell_2$ norm works almost as well as minimizing
the total variation but is much more efficient. If gradients
(edges of anomalies) should be resolved with a minimum of
distortion, we prefer $\ell_1$ damping of Daubechies-4 wavelet
coefficients. It has the additional advantage of yielding a
noiseless reconstruction, contrary to simple $\ell_2$
minimization (`Tikhonov regularization') which should be
avoided. In some of our examples, the $\ell_0$ method produced
notable artifacts. In addition we show how nonlinear $\ell_1$
methods for finding sparse models can be competitive in speed
with the widely used $\ell_2$ methods, certainly under noisy
conditions, so that there is no need to shun $\ell_1$
penalizations.
\end{abstract}

\section{Introduction}

Since geophysical inverse problems are almost always
underdetermined, regularization techniques are essential to
obtain a meaningful solution. Two major classes of techniques
exist. The first one, named `mollifying' in the mathematical
literature, or `optimally localized averaging' (OLA) in
helioseismology, can be traced back to the groundbreaking work
of Backus and Gilbert \cite{backus67,backus70} in geophysics.
In this approach one searches for the size of an averaging
volume that can produce a local average of the model parameter
with an acceptable variance. Since this method is
computationally very expensive, it has found little application
in large-scale geophysical inversions such as seismic
tomography. To limit the computational effort, seismic
tomographers instead search for a biased (`damped') solution.
This has the disadvantage of introducing a systematic error --
the bias -- in lieu of the random error caused by the
propagation of data errors. It can be turned into an advantage
if the bias actually reflects a justified disposition of the
scientist to prefer certain models over others, as long as the
data are fit within their error bars.

Simple $\ell_2$-norm damping, which biases model perturbations
towards zero in the absence of information based on the data,
is generally a bad choice to regularize the inverse problem for
seismic tomography as it tends to introduce structures
reflecting ray coverage into the images. For that reason, most
tomographers prefer to bias the model towards `smooth'
anomalies, in effect trying to forge a compromise between
Backus-Gilbert theory and the efficiency of damped inversions.
The smoothness of the images has the advantage that large
structures become easily visible. Sharp discontinuities,
however, are blurred, and smaller structures, even when
resolved, may be diluted beyond recognition. Recently,
\cite{LoNDD2007} --- hereafter referred to as Paper I ---
introduced a third option for the bias in geophysical
inversions: to minimize the $\ell_1$-norm of the wavelet
decomposition. This also biases the model towards zero, but it
turns out that such reconstructions always have many (or most)
wavelet coefficients \emph{exactly} equal to zero (i.e. they
are sparse). In a synthetic 2D experiment using surface wave
data, we showed how structurally coherent features (in a
geophysical sense), were more faithfully reproduced using this
technique than with a simple $\ell_2$-norm damping. In
addition, as a result of their inherent sparsity, $\ell_1$
reconstructions exhibit much less noise than their $\ell_2$
counterparts.

Though Paper I clearly showed the feasibility of wavelet-based
$\ell_1$ regularization it left a number of questions
unanswered, in particular which wavelet families work best, how
they would perform against more sophisticated $\ell_2$ norms
(e.g. smoothness damping) and whether the computational
feasibility as well as the positive conclusions for wavelet
regularization do scale up to large, 3D models.

In this paper we therefore aim to refine the original
conclusions of Paper~I and to enlarge the scope of the
investigation. We extend tests to 3D inversions of body wave
travel times and investigate the use of different families of
wavelets (Haar, D4, dual tree). We include a comparison with
smoothness damping, and with a fourth option named `total
variation' damping. Unlike Paper~I, we also discuss so-called
$\ell_0$ constrained recovery. In contrast to the $\ell_1$ norm
technique which relies on iterative soft-thresholding, the
$\ell_0$ recovery method uses iterative
\emph{hard}-thresholding of wavelet coefficients
\cite{Blumensath.Davies2008,Blumensath.Davies2009a}.

The (salt dome) model that we try to reconstruct here is more
realistic than the model in Paper~I and includes a wide range
of length scales. The problem described in Paper~I was also of
a very limited size: there were only about $10^4$ degrees of
freedom in the reconstructed models. Here we perform 3D
reconstructions, and increase the number of degrees of freedom
by an order of magnitude to about $\sim10^5$. The number of
data also increases accordingly to $24000$.  Our approach is
complementary to that of \cite{chevrot07} who expand the
Fr\'echet kernels into wavelets to obtain a significant
reduction in the memory requirements to store the kernel.

One disadvantage of the $\ell_1$-norm, $\ell_0$ and total
variation penalties is that they lack the convenient linearity
of the more conventional $\ell_2$-norm minimizations. Making
use of recent algorithmic improvements, we do demonstrate that
finding a (nonlinear) sparse model reconstruction is not
necessarily more expensive, computation-wise, than $\ell_2$
based (linear) reconstructions.

The use of $\ell_1$ norms in seismic tomography was to the best
of our knowledge first proposed in \cite{scales88} in the form
of an iteratively reweighted least squares method (IRLS, see
also \cite{nolet87}), but has never found much favor, possibly
because the convergence of IRLS was not guaranteed. Besides its
use in seismic tomography $\ell_1$ norms have found application
in other geophysical contexts such as deconvolution and
interpolation
\cite{Taylor.Banks.ea1979,Oldenburg.Scheuer.ea1983,Santosa.Symes1986,Sacchi.Ulrych.ea1998}.
The use of $\ell_1$ norms in combination with a carefully
chosen basis (such as wavelets) is, however, more recent and is
largely inspired by the recent development of compressed
sensing \cite{CaRoT2006,Donoho2006,Donoho2006a}, that shows
that under certain conditions an exact reconstruction can be
obtained by solving an $\ell_1$ regularized inverse problem,
provided there is an underlying basis in which the desired
model is sparse. This emphasizes the importance of studying
different ``dictionaries'' as we do here. Recently
\cite{herrmann08crsi} has successfully applied the compressed
sensing idea to wavefield reconstruction, albeit on small-scale
problems only. The success and promise of compressed sensing
has therefore also increased interest in the speed-up of such
$\ell_1$ problems to be able to handle practical geophysical
applications (e.g.
\cite{Hennenfent.Berg.ea2008,Berg.Friedlander2008}).

\section{Forward problem formulation}

\label{setupsection}

We plan to test the regularization methods on a synthetic data
set generated for a salt dome model. Since the main goal of
this paper is to evaluate and compare a number of algorithms,
numerical efficiency is more important than the wish to have a
tomographic problem at hand that is fully realistic. We have
thus taken a few shortcuts to be able to run inversions quickly
in Matlab on a single processor. However, we took pains to
ensure that we would invert for a model that has a large range
of length scales, and that the ray coverage would encompass
both dense and sparse regions.

In dimensionless variables, the expression for a finite
frequency sensitivity kernel corresponding to a constant
background and a Gaussian power spectrum is given by the
formula:
\begin{equation}
K(x,y,z)=\frac{\displaystyle e^{-u^2}H_5(u)}{24\lambda d_\mathrm{s} d_\mathrm{r}}\, ,
\label{kernel}
\end{equation}
where $u=\pi
(d_\mathrm{s}+d_\mathrm{r}-d_\mathrm{sr})/\lambda$. Here
$d_\mathrm{sr}$ is the distance between source (earthquake) and
receiver (station), and $d_\mathrm{s},d_\mathrm{r}$ are the
distances to source and to receiver, measured from the point
$(x,y,z)$. $\lambda$ is the dominant wavelength and
$H_5(u)=120u-160 u^3+32u^5$ denotes the Hermite polynomial of
order 5. Equation (\ref{kernel}) can be derived from the
expressions for the Fr\'echet kernel in a homogeneous medium
using an analytical expression for the spectral integration
(see \cite{Favier.Chevrot2003}).


The constant background model that we use here is not at all
realistic from a physical point of view for the salt dome input
model that we will use in section \ref{reconstructionsection}:
in case of of large velocity contrasts kernels are bent rather
than straight as in expression (\ref{kernel}). However, as we
will use the same constant background kernels (\ref{kernel})
for generating synthetic data as well as for reconstructing
models from these data, we believe it is possible to accurately
evaluate the effects of the regularization technique on the
inversions. Obviously, reconstructing a model from actual
measurements requires kernels of a more complicated shape than
(\ref{kernel}) but their resolving power is not fundamentally
different from those used in our simple application.

For each source-receiver pair and each dominant wavelength, the
travel time differential and the model perturbation $\m$ are
connected by the linear integral relation
\cite{Nolet2008}\footnote{For convenience, we denote the model
perturbation by $\m$ instead of $\delta\m$ as is often done.}
\begin{equation}
\delta T=\int_V \, K\, \m\, \mathrm{d}V \,.
\end{equation}
Given sufficiently many data, the aim of seismic tomography is
to reconstruct the model $\m$ from a noisy version of the data
vector $\data$ containing many travel time differentials
corresponding to many source-receiver pairs.

In section \ref{reconstructionsection} we will perform a number
of seismic reconstructions. The domain on which we will do this
is the cube $V=[-1, 1]^3$. For discretization, this domain is
subdivided in $64^3$ voxels, a convenient choice for the
digital reconstruction of a model $\m$, leading to $262144$
degrees of freedom in $\m$. In order to be able to produce
meaningful reconstructions, we expect to need at least $10^4$
data (about $1$ datum for $10$ degrees of freedom). Hence we
will choose $4800$ sources-receiver pairs and $5$ different
dominant wavelengths so as to yield $24000$ data. This
represents an overparametrization by a factor of more than 10.
Thus regularization will be an essential requirement for any
data inversion.

\begin{figure}
\centering
\includegraphics{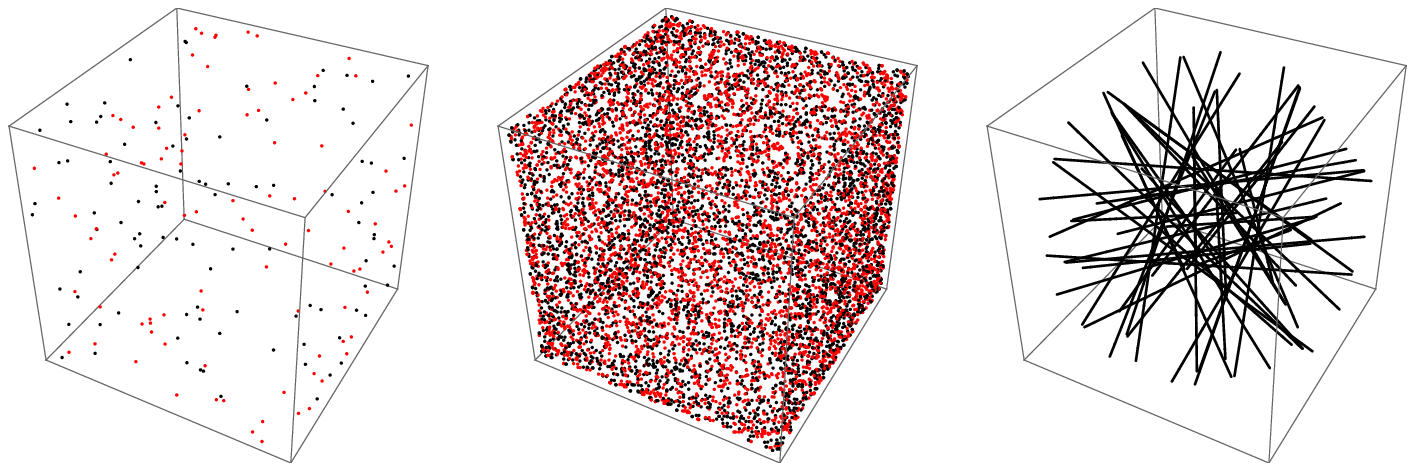}
\caption{Left: The cube $[-1,1]^3$ with the initial $100$ source-receiver pairs (black=source, red=receiver). Center: The $4800$ source-receiver pairs
one obtains by applying the $48$ symmetry transformations of Table \ref{transformationtable} on the initial $100$ pairs.
Right: A single source-receiver pair joined by a straight line and its $48$ transformations.
Plotting all $4800$ rays obtained in this way would fill the whole cube.}\label{transformationpic}
\end{figure}

A very efficient set-up of our numerical experiment was
obtained as follows: we first choose $100$ source-receiver
pairs in random positions on the surface of the cube, while
making sure source and receiver are never on the same face (the
kernels (\ref{kernel}) are not curved and would not be able to
cross the model domain very much if source and receiver were on
the same face). From these initial $100$ pairs, we construct
the full set of $4800$ pairs by using the $48$ symmetry
transformations of the cube. These $48$ operations are
constructed from the $3!$ permutations of the coordinates
$(x,y,z)$, and by the $2^3$ reflections $(\pm x,\pm y,\pm z)$,
as listed in Table \ref{transformationtable}. In other words,
starting from one initial kernel $K(x,y,z)$ we easily obtain
$47$ other kernels: $K(z,x,y)$, $K(y,x,-z)$, \ldots,
corresponding to differently positioned source-receiver pairs.
The random nature of the positions of the original $100$
source-receiver pairs (i.e. no coordinate is exactly zero or
exactly equal to plus or minus another coordinate) ensures us
that none of the $4800$ source-receiver pairs are identical.
The initial $100$ source-receiver pairs and the final $4800$
pairs are shown in Figure \ref{transformationpic}.

\begin{table}
\centering $\begin{array}{l|llllll} &
\multicolumn{6}{c}{\mathrm{Permutations}}\\\hline
\multirow{8}{*}{\rotatebox{90}{Sign changes}} & (x,y,z) & (x,z,y) & (y,x,z) & (y,z,x) & (z,x,y) & (z,y,x) \\
& (x,y,-z) & (x,z,-y) & (y,x,-z) & (y,z,-x) & (z,x,-y) & (z,y,-x) \\
& (x,-y,z) & (x,-z,y) & (y,-x,z) & (y,-z,x) & (z,-x,y) & (z,-y,x) \\
& (x,-y,-z) & (x,-z,-y) & (y,-x,-z) & (y,-z,-x) & (z,-x,-y) & (z,-y,-x) \\
& (-x,y,z) & (-x,z,y) & (-y,x,z) & (-y,z,x) & (-z,x,y) & (-z,y,x) \\
& (-x,y,-z) & (-x,z,-y) & (-y,x,-z) & (-y,z,-x) & (-z,x,-y) & (-z,y,-x) \\
& (-x,-y,z) & (-x,-z,y) & (-y,-x,z) & (-y,-z,x) & (-z,-x,y) & (-z,-y,x) \\
& (-x,-y,-z) & (-x,-z,-y) & (-y,-x,-z) & (-y,-z,-x) &
(-z,-x,-y) & (-z,-y,-x)
\end{array}$
\caption{Left: A list of rotations and reflections that map the
unit cube $[-1,1]^3$ onto itself. They are constructed by
combining the six permutations of $(x,y,z)$ and eight sign
changes of the components
$(x,y,z)$.}\label{transformationtable}
\end{table}

For each of the $4800$ source-receiver pairs we will construct
five finite frequency sensitivity kernels of type
(\ref{kernel}) corresponding to five different dominant
wavelengths $\lambda=\{0.5, 0.2, 0.08, 0.04, 0.025\}$. Thus, in
total there will be $24000$ kernels at our disposal. Because of
the symmetry transformations used and the random choice of the
initial 100 source-receiver pairs, the coverage of the domain
by these 24000 kernels is quite uniform. A picture that
illustrates this property is too large to include here, but it
is available in the online supplementary material in Figure
A.1.

\begin{figure}
\resizebox{\textwidth}{!}{\includegraphics{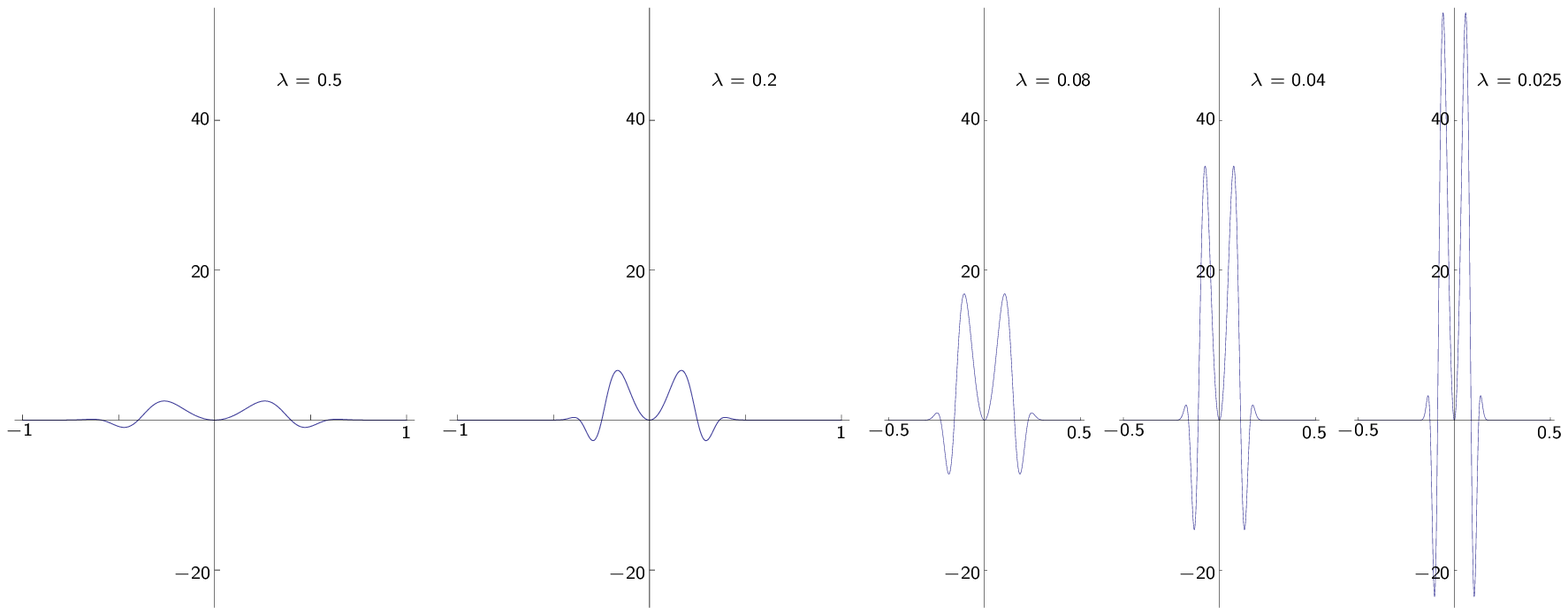}}
\caption{Cross sections (perpendicular to the midpoint of the central ray) of five finite sensitivity kernels corresponding to
$\lambda\in\{0.5, 0.2, 0.08, 0.04, 0.025\}$. Larger wavelength kernels are wider but have smaller amplitude.}\label{kernelpic}
\end{figure}

In order to give the reader an idea of the size of the Fresnel
zones of these finite frequency kernels, we include in
Figure~\ref{kernelpic} and in one of the panels of
Figures~\ref{saltdomepic1blowup} and \ref{saltdomepic2blowup} a
cross-sectional view of five kernels. Each of these kernels
corresponds to one of the five wavelengths ($\lambda\in\{0.5,
0.2, 0.08, 0.04, 0.025\}$) and source-receiver distance
$d_\mathrm{sr}=2$. The width of the Fresnel zone is
proportional to $\sqrt{\lambda d_\mathrm{sr}}$
\cite{Dahlen2004}. The cross sections in Fig.~\ref{kernelpic}
illustrates not only the typical widths of the kernels, but
also their relative amplitudes.

Additionally we construct a second operator containing only
$20000$ out of the $24000$ kernels. We choose to remove the
$4000$ kernels for which the line connecting source and
receiver (i.e. the central ray) comes closest to the point
$(0.24, -0.7, -0.23)$. In this way, we end up with an operator
that has a `hole' in its coverage of the cube (see Figure A.1,
right). Using this operator we will be able to study the effect
of non-uniform coverage on reconstructions.

To discretize the model into voxels we calculate the integral
of the sensitivity kernels over each voxel (using a Riemann sum
with $4^3$ terms/voxel) for each source-receiver pair and for
each of the five dominant wavelengths we consider. The
resulting values make up the operator $\A$ we want to invert:
Because there are $24000$ kernels and $64^3=262144$ voxels, the
matrix $\A$ will have $24000$ rows and $262144$ columns.

The use of the $48$ symmetry transformations of the cube allows
us to save a factor of $48$ in memory requirements for our
calculations, i.e. we  need to compute and store only $500$
kernels corresponding to the initial $100$ source-receiver
pairs and $5$ wavelengths. The remaining ones are easily (and
quickly) generated from these $500$ kernels by the symmetry
transformations of the cube (quick rearrangements of the
elements in a 3D array).

An additional saving in memory requirements is obtained by
exploiting the fact that most kernels are well localized (i.e.
they are thin), and thus that the $A_{ij}$ are practically zero
for many voxels. In other words, each row of the matrix $\A$ is
relatively sparse.

Although this set-up is completely unrealistic from a physical
perspective, this configuration of rays provides for an easy
way to compare dense and partial data coverage of the model. It
also allows us to focus on the relative merits of the inversion
methods rather than on the difficulties a physically faithful
modeling would entail.

In reality, sensor coverage is limited to the surface and a few
boreholes. High frequency data with narrow Fresnel zones
therefore leave significant areas at depth not illuminated by
acoustic waves. Low frequency data have wider Fresnel zones but
suffer a reduced sensitivity to small length scales.
Illumination itself does not guarantee resolution as becomes
clear when one regards the singular value spectrum of $\A$:
though the spectrum depends strongly on the experimental
set-up, it often has a rapid drop-off and a significant
fraction of eigenvalues is either zero or too small to be
useful. In the present synthetic experiment, we chose to have
about 1 datum for every 10 degrees of freedom in the model. The
uniform random distribution of sources and receivers on the
surface of the cube has a favorable influence on the singular
value spectrum of $\A$ (we were able to confirm this on a
scaled-downed version of the present operator $\A$, but not on
the full $24000\times 262144$ matrix), and will aid in the
reconstruction . This will allow us to focus on the
characteristics of the different reconstruction methods, rather
than on the lack of data.

\section{Reconstruction methods} \label{algorithmsection}

Reconstructing the model $\m$ from the data $\data$ is done, in
principle, by solving the linear system $\A\m=\data$, where
$\A$ denotes, as before, the matrix containing the kernels
discretized in the voxel basis. This system may contain
incompatible equations (due to noise), and at the same time be
underdetermined (not enough data to reconstruct all of $\m$).

The problem of incompatible data can be solved by replacing the
original problem with the minimization of a data fidelity term:
\begin{equation}
\bar\m=\arg\min_\m \|\A\m-\data\|^2\,.\label{datamisfit}
\end{equation}
Here and in the following $\|u\|$ (without subscripts) always
denotes the usual 2-norm of u: $\|u\|=\sqrt{\sum_iu_i^2}$.
Although a minimizer always exists (because of the quadratic
nature of the functional), it may not always be unique. In
other words, the problem is still underdetermined and an
iterative numerical scheme for finding a minimizer of
(\ref{datamisfit}) may diverge. In fact, because of the
existence of data errors we are not even looking for the exact
minimizer (\ref{datamisfit}). We rather augment the functional
in (\ref{datamisfit}) by a term that will penalize a whole
category of models that is thought to be unphysical.

\subsection{$\ell_2$ penalties}

The prime example of this kind of method is Tikhonov
regularization \cite{Tikhonov1963} whereby a penalty
proportional to the $\ell_2$-norm of the model is imposed:
\begin{equation}
\bar \m=\arg\min_\m\|\A \m-\data\|^2+\mu
\|\m\|^2\,.\label{l2functional}
\end{equation}
This will effectively prevent the model from growing
unboundedly due to noise in $\data$ and ill-conditioning of the
matrix $\A$.

Another, closely related, possibility is to impose a penalty
consisting of the $\ell_2$-norm of the (discrete) Laplacian
$\Delta\m$ of $\m$:
\begin{equation}
\bar \m=\arg\min_m\|\A \m-\data\|^2+\mu
\|\Delta\m\|^2\,.\label{l2deltafunctional}
\end{equation}
As we will see in section \ref{reconstructionsection}, this
will enforce a certain degree of smoothness on the
reconstructed model. The Laplacian used here is defined as the
difference of the model with the local average over the six
nearest neighbors:
$(\Delta\m)_{ijk}=\m_{ijk}-(\m_{i-1jk}+\m_{i+1jk}+\ldots+\m_{ijk+1})/6$.

The above two versions of the Tikhonov regularization method
have the advantage of being solvable by linear equations. The
variational equations that determine the minimizers of
(\ref{l2functional}) and (\ref{l2deltafunctional}) are:
\begin{equation}
\A^T\A\m+\mu\,\m=\A^T\data\,,
\end{equation}
and
\begin{equation}
\A^T\A\m+\mu\,\Delta^T\Delta\m=\A^T\data\,,
\end{equation}
respectively (with suitable treatment of boundary voxels,
$\Delta^T=\Delta$ in the latter case). For these linear
equations, we can use the conjugate gradient algorithm. With
$\m^{(0)}=$arbitrary and $\rr^{(n)}=\A^T(\data-\A\m^{(n)})-\mu
\D^T\D\m^{(n)}$, we set
\begin{equation}
\begin{array}{l}
\displaystyle \vv^{(n)}=\left\{\begin{array}{ll}
\rr^{(n)} & n=0\\
\rr^{(n)}+\frac{\|\rr^{(n)}\|^2}{\|\rr^{(n-1)}\|^2} \vv^{(n-1)} & n>0
\end{array}\right.\\
\displaystyle \m^{(n+1)}=\m^{(n)}+\frac{\|\vv^{(n)}\|^2}{\|\A \vv^{(n)}\|^2+\mu \|\D \vv^{(n)}\|^2} \vv^{(n)},
\end{array}\label{conjgrad}
\end{equation}
where $\D$ is either the unit matrix $\mathbf{I}$ (in case we
seek to minimize functional (\ref{l2functional})) or $\Delta$
(in case we use functional (\ref{l2deltafunctional})). The
model estimates $\m^{(n)}$ converge to the minimizer
(\ref{l2functional}) or (\ref{l2deltafunctional}), respectively
as $n$ increases.

\subsection{$\ell_1$ penalties}

Another and much more recent method of regularization consists
of imposing a carefully chosen $\ell_1$-norm penalty
\cite{Daubechies2004b}. It can be shown that this leads to a
sparse model, i.e. a model with few nonzero components
\cite{Donoho2006,Donoho2006a}. It would therefore \emph{not} be
a good idea to apply this technique to the model in the voxel
basis (there is no reason to assume the model would be sparse
in that basis); we would rather use this penalty on the
coefficients of the model in a different basis in which we
believe the model to be sparse.

Harmonic functions would allow us to select resolvable scales,
but the complete lack of localization of these functions makes
them even worse candidates than voxels. 3D wavelets offer a
compromise between the concentration of power in both scale and
location, and are intuitively more suitable to build
geophysically reasonable models. In fact, our earlier
experience (Paper I) showed the advantages of using a wavelet
basis, and constructing the model by finding the minimum of the
functional
\begin{equation}
\|\A\m-\data\|^2+2\mu \|\w\|_1,
\end{equation}
where $\w=W\m$ are the wavelet coefficients of $\m$. This
minimization problem can be rewritten as:
\begin{equation}
\bar \m=\W^{-1}\bar\w\quad\mathrm{and}\quad\bar\w=\arg\min_\w F(\w)\quad \mathrm{with}\quad
F(\w)=\|\A
\W^{-1}\w-\data\|^2+2\mu \|\w\|_1. \label{l1functional}
\end{equation}
$\W$ is the wavelet decomposition matrix and $\W^{-1}$ the
wavelet synthesis operator. This type of $\ell_1$-norm penalty
leads to a model that has a sparse wavelet representation, i.e.
a model with (very) few nonzero wavelet coefficients. The aim
is thus to rely on the properties of the wavelet basis to be
able to represent the desired solution with few nonzero
components. In geophysics wavelets are a good choice for
seismic reconstruction as they allow for sparse representations
of overall smooth functions, while still capable of taking into
account the possibility of isolated sharp features
\cite{Daubechies1992,LoNDD2007}.

Another advantage of the $\ell_1$ method is that they yield
noise-free model reconstructions. This is a consequence of the
simple fact that noise in the model cannot be represented in a
sparse way (in any reasonable basis). This method is thus able
to produce clean models without the need for additional
smoothing. It is important to note that the least squares
functional (\ref{l1functional}) is convex. This implies that a
local minimum of (\ref{l1functional}) is always a global
minimum as well.

In section \ref{reconstructionsection}, we shall consider a
number of different choices of orthonormal wavelet bases. For
each of these choices, we have $\W^{-1}=\W^T$, which we will
implicitly assume hereafter.

In order to find the minimizer $\bar \w$ of the $\ell_1$
penalized functional (\ref{l1functional}), one may use the
iterative soft-thresholding algorithm \cite{Daubechies2004b}:
\begin{equation}
\w^{(n+1)}=T(\w^{(n)}),
\label{ist}
\end{equation}
with
\begin{equation}
T(\w)=\mathcal{S}_{\alpha\mu}\left[\w+\alpha \W\A^T (\data-\A\W^T\w)\right],
\label{Top}
\end{equation}
and where $\w^{(0)}$ may be chosen arbitrarily. The
soft-thresholding $\mathcal{S}_\tau(u)$ function operates
component-wise and is defined by
\begin{equation}
\mathcal{S}_\tau(u)=
\left\{\begin{array}{llccc}
u-\tau & & u  & \geq & \tau\\
0      & & |u|& \leq & \tau\\
u+\tau & & u  & \leq & -\tau.
\end{array}\right.
\end{equation}
This algorithm was used in a 2D seismic tomography toy problem
in Paper I, to which we also refer for a brief but elementary
derivation (section 2 of paper I). In effect, it is a simple
gradient descent algorithm (with fixed step length $\alpha$)
where the additional soft-thresholding operation
$\mathcal{S}_{\alpha\mu}$ is a mathematical consequence of the
$\ell_1$-norm term present in functional (\ref{l1functional}).

The constant $\alpha$ should be chosen such that $\alpha
\|\A^T\A\|$ is smaller than or equal to $1$ ($\|\A^T\A\|$ is
defined as the largest eigenvalue of
$\A^T\A$)\cite{Daubechies2004b}. In our calculation we always
choose $\alpha=0.95/\|A^TA\|$.

The wavelet transform $\W$ and its inverse $\W^T$ are fast
transforms. This means that they cost only a fraction of the
computer time needed to perform one application of $\A$ or
$\A^T$. In other words, working in a wavelet basis does not
significantly change the computational complexity of the
reconstruction algorithm.


As the iterative soft-thresholding algorithm (\ref{ist}) can be
slow in practice, we have opted here for using the so-called
Fast Iterative Soft-Thresholding Algorithm (FISTA)
\cite{Beck.Teboulle2008} (see also earlier work of Nesterov
\cite{Nesterov1983,Nesterov1983a}):
\begin{equation}
\w^{(n+1)}=T\left(\w^{(n)}+\frac{t_n-1}{t_{n+1}} \left(\w^{(n)}-\w^{(n-1)}\right)\right),
\label{fista}
\end{equation}
with $t_0=1$ and $t_{n+1}=(1+\sqrt{1+4t_n^2})/2$.\footnote{In
the journal published version of this manuscript this formula
contains a typo.} The FISTA algorithm has practically the same
computational complexity as the iterative soft-thresholding
algorithm (\ref{ist}). It only requires one additional vector
addition. With this algorithm the $\ell_1$ penalized cost
function (\ref{l1functional}), evaluated at $\w=\w^{(n)}$, is
bounded by $\mathcal{O}(1/n^2)$ from its limiting value:
\begin{equation}
F(\w^{(n)})-F(\bar \w)\leq 4\frac{\|\w^{(0)}-\bar \w\|^2}{\alpha\, (n+1)^2},\label{fistabound}
\end{equation}
as opposed to the $\mathcal{O}(1/n)$ decrease:
\begin{equation}
F(\w^{(n)})-F(\bar \w)\leq \frac{\|\w^{(0)}-\bar \w\|^2}{\alpha\, n},
\end{equation}
that can be proven for algorithm (\ref{ist}). These upper
bounds are valid non-asymptotically, i.e. even for small $n$
\cite{Beck.Teboulle2008,Nesterov2004}.

\subsection{$\ell_0$ penalties}

Recently, mathematical advances on direct ways of constraining
the number of nonzero components in a reconstructed model have
appeared. In
\cite{Blumensath.Davies2008,Blumensath.Davies2009a} an
iterative hard-thresholding algorithm is proposed of the
following form:
\begin{equation}
\w^{(n+1)}=\tilde T(\w^{(n)}),
\label{iht}
\end{equation}
with
\begin{equation}
\tilde T(\w)=\mathcal{H}_{k}\left[\w+\alpha \W\A^T (\data-\A\W^T\w)\right]
\label{tildeTop}
\end{equation}
and where the hard-thresholding operation
$\mathcal{H}_{k}(\uu)$ sets all but the largest (in absolute
value) $k$ components of $\uu$ to zero. This algorithm
converges to a \emph{local} minimum of
\begin{equation}
\|\A\W^{-1}\w-\data\|^2\qquad\mathrm{under\ the\ condition}\qquad \|\w\|_0\leq k.\label{l0functional}
\end{equation}
Here $\|\w\|_0$ denotes the number of nonzero coefficients of
$\w$. Just as with the $\ell_1$ penalty method, we shall apply
this technique using a wavelet basis. The underlying reason is
again the suitability of a wavelet basis to represent a
physically acceptable model with few nonzero wavelet
coefficients. We stress that the appearance of the
hard-thresholding operation $\mathcal{H}_k$ is a mathematical
consequence of the constraint $\|\w\|_0\leq k$, just as
soft-thresholding in (\ref{Top}) is a mathematical consequence
of the $\ell_1$ term in (\ref{l1functional})
\cite{Daubechies2004b,LoNDD2007}.

The algorithm (\ref{iht}) converges very slowly. We therefore
propose the following FISTA/Neste\-rov-like modification:
\begin{equation}
\w^{(n+1)}=\tilde T\left(\w^{(n)}+\frac{t_n-1}{t_{n+1}} \left(\w^{(n)}-\w^{(n-1)}\right)\right),
\label{fiht}
\end{equation}
with $t_0=1$ and $t_{n+1}=(1+\sqrt{1+4t_n^2})/2$.\footnote{In
the journal published version of this manuscript this formula
contains a typo.} As far as the authors know, this is the first
time this algorithm is proposed. Although there is no proof of
convergence yet, we found that it worked quite well on the
examples that we studied (see section
\ref{reconstructionsection}). We used the same choice for the
step-length $\alpha$ as with the $\ell_1$ algorithm:
$\alpha=0.95/\|A^TA\|$. The choice of the number of nonzero
model wavelet coefficients $k$ in method (\ref{l0functional})
is discussed in section \ref{choicesection}.

On a side note, it would also be possible to calculate a local
minimizer of the (non-convex) functional
\begin{equation}
\|\A\W^{-1}\w-\data\|^2+\mu \|\w\|_0
\end{equation}
using component-wise hard-thresholding (with a fixed threshold
$\sqrt{\mu}$) \cite{Blumensath.Davies2008}. However,
\cite{Blumensath.Davies2009a} seems to prefer the formulation
(\ref{l0functional}) that imposes a \emph{fixed} number of
nonzeros in each step of the iteration, rather than a fixed
lower bound $\sqrt{\mu}$ for the absolute values of the wavelet
coefficients.

Although no proof of convergence of algorithm (\ref{fiht}) is
given, we shall refer to it by the name `$\ell_0$ method'
(instead of calling it `hard-thresholded Nesterov accelerated
gradient descent algorithm').

\subsection{Total variation penalty}

A final penalty term we will consider is the so-called `total
variation' (TV) penalty:
\begin{equation}
\bar \m=\arg\min_\m\|\A\m-\data\|^2+2\mu \sum_{ijk}
\sqrt{(\Delta_x\m)^2+(\Delta_y\m)^2+(\Delta_z\m)^2},
\label{TVfunctional}
\end{equation}
with $\Delta_x \m=\m_{ijk}-\m_{i-1jk}$ and similarly for
$\Delta_y$ and $\Delta_z$. This penalization will favor
piece-wise constant models in the voxel basis. Unfortunately,
the equations that determine the minimizer (\ref{TVfunctional})
are again nonlinear. We will use a reweighed conjugate gradient
method \cite{Oliveira.Bioucas-Dias.ea2009} to determine the
minimum of the TV functional (\ref{TVfunctional}). More
specifically, defining the weights
$\uu^{(n)}=\left(\Delta_x\m^{(n)}+\Delta_y\m^{(n)}+\Delta_z\m^{(n)}\right)^{-1/2}$,
we shall use algorithm (\ref{conjgrad}) where we choose
$\D^T\D=\Delta_x^T \uu^{(n)}\Delta_x+\Delta_y^T
\uu^{(n)}\Delta_y+\Delta_z^T \uu^{(n)}\Delta_z$ (which depends
on the iteration $n$) and use that $\|\D\vv^{(n)}\|=\langle
\D^T\D\vv^{(n)},\vv^{(n)} \rangle$. Because of the
non-quadratic nature of functional (\ref{TVfunctional}), the
conjugate gradient algorithm no longer preserves conjugacy
between successive search directions $\vv^{(n)}$ as $n$ grows.
For this reason, the iteration is also reset every so often (in
accordance with \cite{Oliveira.Bioucas-Dias.ea2009}).

An algorithm similar to (\ref{ist}) and (\ref{fista}) may also
be used to find the minimizer of the TV penalized functional
(\ref{TVfunctional})
\cite{Bect.Blanc-Feraud.ea2004,Bioucas-Dias.Figueiredo2007}. In
fact, in \cite{Bect.Blanc-Feraud.ea2004} it is shown how the
$\ell_1$ norm penalty and the TV penalty can be combined in a
single functional.

\subsection{Choice of the penalty parameter}

\label{choicesection}

As such, the minimizers defined by (\ref{l2functional}),
(\ref{l2deltafunctional}), (\ref{l1functional}) and
(\ref{TVfunctional}) still depend on the penalty parameter
$\mu$, or in case of the hard-thresholding algorithm
(\ref{fiht}) on the parameter $k$. In the reconstructions
below, we select this parameter by requiring in each instance
that the reconstructed model $\bar\m$ fits the data $\data$ as
well as possible, but not any better than the noise level:
\begin{equation}
\|\A\bar \m-\data\|\approx\|\n\|,
\end{equation}
with $\n$ representing the noise vector. In other words, the
discrepancy principle tells us to choose the penalty parameter
$\mu$  (or $k$) such that
\begin{equation}
\|\A\bar \m-\data\|^2/\sigma^2\approx\mathrm{number\ of\ data},
\label{penaltychoice}
\end{equation}
where $\sigma$ is the noise variance (if different data have
different variance, it is simplest to divide from the outset
each row as well as the right hand side by the standard
deviation of its datum, and set $\sigma =1$ in
(\ref{penaltychoice})). In practice, this means that we will
have to perform the minimization several times, until a
suitable value of $\mu$ or $k$ is determined for each
reconstruction.

\section{Reconstructions}
\label{reconstructionsection}

In this section we present some sample reconstructions using
the algorithms mentioned in section \ref{algorithmsection},
applied to the finite frequency tomography problem described in
section \ref{setupsection}. First we will consider a simple
checkerboard input model that we will try to reconstruct from
incomplete and noisy data. We will also look at a more
complicated salt dome model which we obtained from BP America,
Inc.

In each case, our procedure will be the following. We start
from a known input model $\m^\mathrm{input}$ from which we
construct synthetic noisy data $\data$:
\begin{equation}
\data=\A\m^\mathrm{input}+\n.
\end{equation}
The noise $\n$ is taken from a Gaussian distribution, with zero
mean and variance $\sigma$ chosen in such a way that
$\|\n\|/\|A\m^\mathrm{input}\|=0.1$; in other words, we add
$10\%$ noise to the noiseless data. The goal is then to try to
reconstruct $\m^\mathrm{input}$ as well as possible from
$\data$ and $\A$. For this we will use methods
(\ref{l2functional}), (\ref{l2deltafunctional}),
(\ref{l1functional}), (\ref{l0functional}) and
(\ref{TVfunctional}), and compare the results. Since we know
the noise variance $\sigma$, we can use the criterion
(\ref{penaltychoice}) to choose the penalty parameter $\mu$ or
$k$. This parameter will be different for the various synthetic
data and for the various penalties that we impose.

For the wavelet based methods (\ref{l1functional}) and
(\ref{l0functional}), we also need to choose a specific wavelet
family. There are many wavelet bases, with varying degrees of
smoothness and approximation properties \cite{Daubechies1992}.
This gives us the opportunity to adapt our choice of wavelet
basis to the model: we will choose the basis in which we
suspect the model to be sparse. In particular, for the
checkerboard reconstruction we will use Haar wavelets because
we know that the true solution is very sparse in that basis. We
thus expect a very accurate reconstruction in that case. For
the salt dome input model, we will compare the effects of
different choices of wavelets bases on the reconstruction. Our
choice will include Haar wavelets, D4 wavelets and also
directional dual tree wavelets. Haar wavelets are the least
smooth, and directional wavelets are the most smooth of these
three.

\subsection{Checkerboard}
\label{checksection}

The first example consists of a checkerboard pattern; in other
words the input model $\m^\mathrm{input}$ is piecewise constant
($\pm 1$) on small cubes of $8$ by $8$ by $8$ voxels. It mainly
serves as a proof of principle for the $\ell_1$ wavelet method
(\ref{l1functional}) because we know that this input model is
sparse in the Haar wavelet basis \cite{Haar1910}. The Haar
wavelets are piecewise constant. In fact, the model only has
$99$ nonzero coefficients in that basis (out of $262144$).
Hence, we may expect that the $\ell_1$ method will work very
well with this model and this basis.

\begin{figure}
\centering\includegraphics{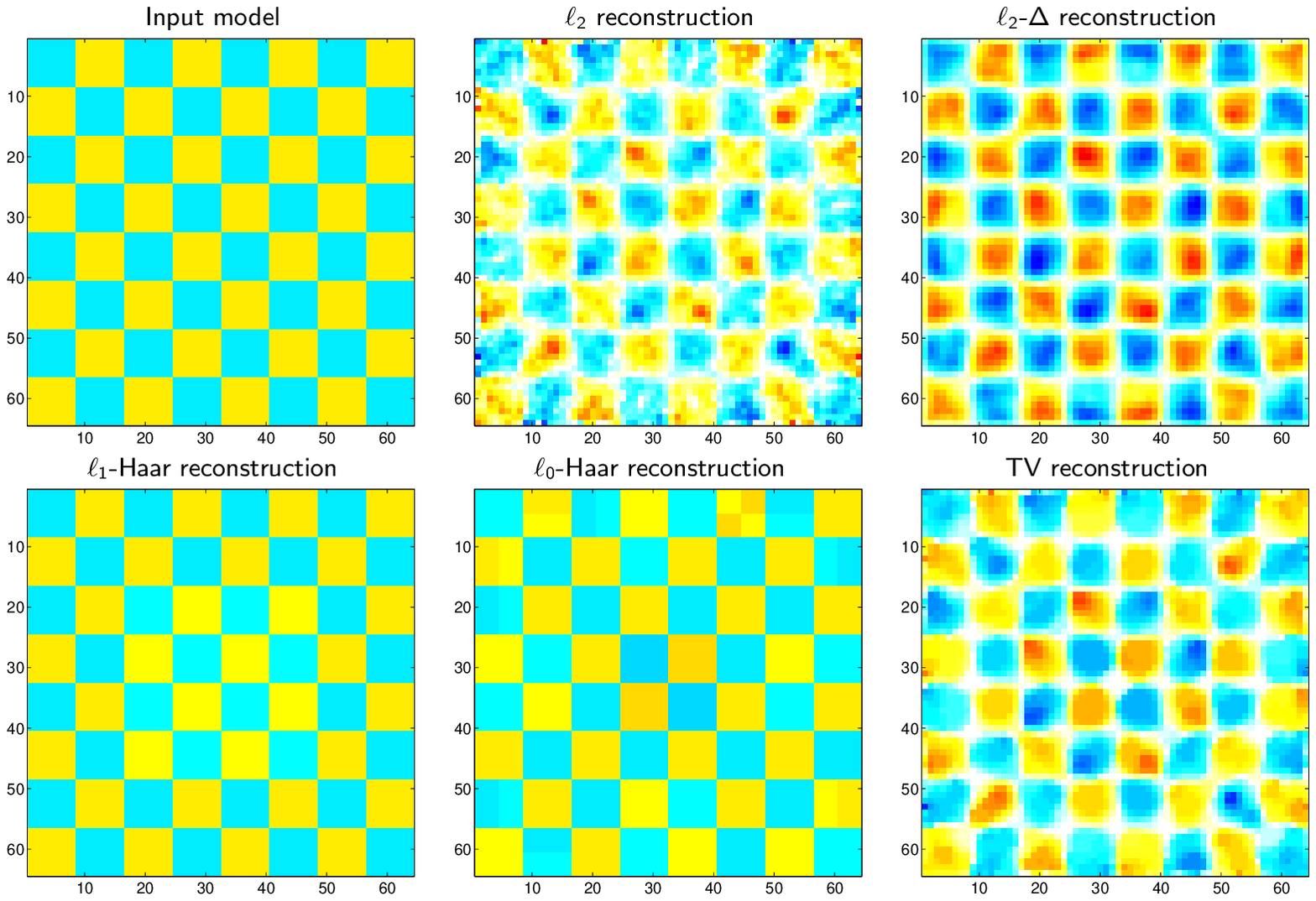}
\caption{A single horizontal slice (number 25 (from top) out of 64) of the checkerboard model and its reconstructions using all
$24000$ data. The whitish transitions in several of the reconstructions
are the result of smoothing between $+1$ and $-1$ of the input model.
As the input checkerboard has fields of size $8\times8\times8$
with constant model value $\pm1$, the smoothing occurs with
period $8$ as well. In addition, the $\ell_2$ reconstruction has a distinctive noisy appearance.
The $\ell_1$ reconstruction using Haar wavelets is visually indistinguishable from the input model. The $\ell_0$ method does almost as well.}
\label{checkerpic1}
\end{figure}

A single horizontal slice of the checkerboard input model and
its four reconstructions are shown in Figure \ref{checkerpic1}.
These four reconstructions use all $24000$ kernels so that no
particular region in the model is favored or disadvantaged. The
$\ell_1$ reconstruction (\ref{l1functional}) with Haar wavelets
is visually the most faithful to the original, closely followed
by the $\ell_0$-Haar reconstruction. The
$\ell_2$-reconstructions
(\ref{l2functional},\ref{l2deltafunctional}) and the TV
reconstruction (\ref{TVfunctional}) display smooth transitions
between $+1$ and $-1$. The reconstruction result of the simple
$\ell_2$ method (\ref{l2functional}) is quite noisy, which is
not the case for the other methods. For completeness and
viewing convenience, the online supplementary material includes
a picture (Figure A.2) of all the horizontal slices of the
input model and its various reconstructions.

\begin{figure}\centering
\resizebox{\textwidth}{!}{\includegraphics{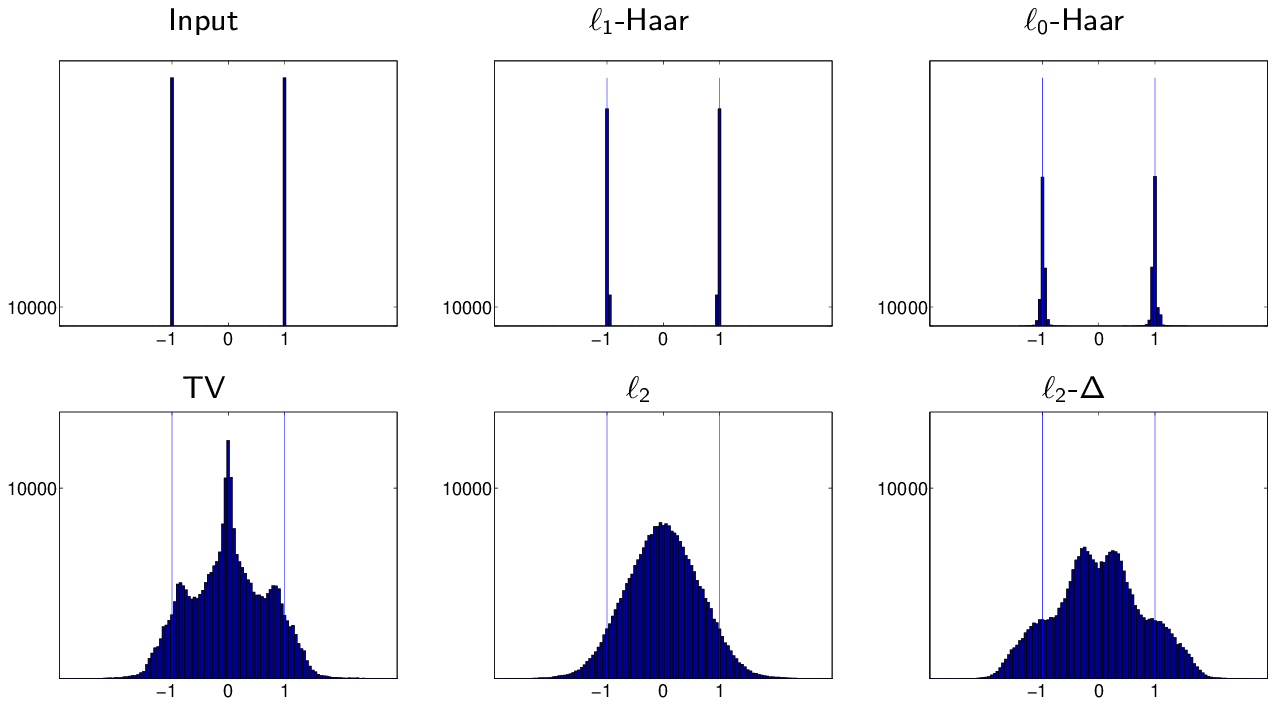}}
\caption{Histograms of the amplitude distribution of the five
checkerboard reconstructions. The input model takes on values $+1$ and $-1$ only. The $\ell_1$ reconstruction in
the Haar basis, and to a lesser extent the $\ell_0$-Haar reconstruction, result in almost
perfect reconstruction of the amplitude distribution
whereas the amplitudes are
shifted towards zero by the other reconstruction methods.} \label{histpic}
\end{figure}

The smoothing effect of the reconstructions can quantitatively
be seen from the histogram of the reconstructed model
amplitudes (see Figure \ref{histpic}). The $\ell_2$
reconstruction takes on mostly values around zero, whereas the
input model only has amplitudes $+1$ and $-1$ (vertical blue
lines). In this case, the $\ell_1$ Haar reconstruction does a
very good job at recovering the $\pm 1$ amplitude distribution,
as it is naturally well suited for the particular checkerboard
model used. The $\ell_0$ method does second best and the TV
method (\ref{TVfunctional}) does third best from this point of
view. It is surprising that the $\ell_1$ method outperforms the
$\ell_0$ method in this case.

The checkerboard model shows that good reconstructions are
possible if one has very good prior knowledge on the model.
When using the  $\ell_0$ and $\ell_1$ methods this requires a
basis in which the desired model is very sparse. The
checkerboard model that was used here aligns optimally with the
chosen Haar basis. The results of $\ell_1$ and $\ell_0$
reconstructions deteriorate when the checkerboard pattern is
shifted w.r.t the Haar basis or when the size of the fields are
changed (the Haar basis decomposition of such a checkerboard
would seize to be sparse). As such, the checkerboard model we
choose is very particular. For realistic reconstruction
scenarios one should not expect such excellent results.

A discussion of the mean square error of the various
reconstructions is given in section \ref{compaspectssection},
where other computational aspects are also discussed.

\subsection{A 3D salt dome model}

\begin{figure}
\centering\includegraphics{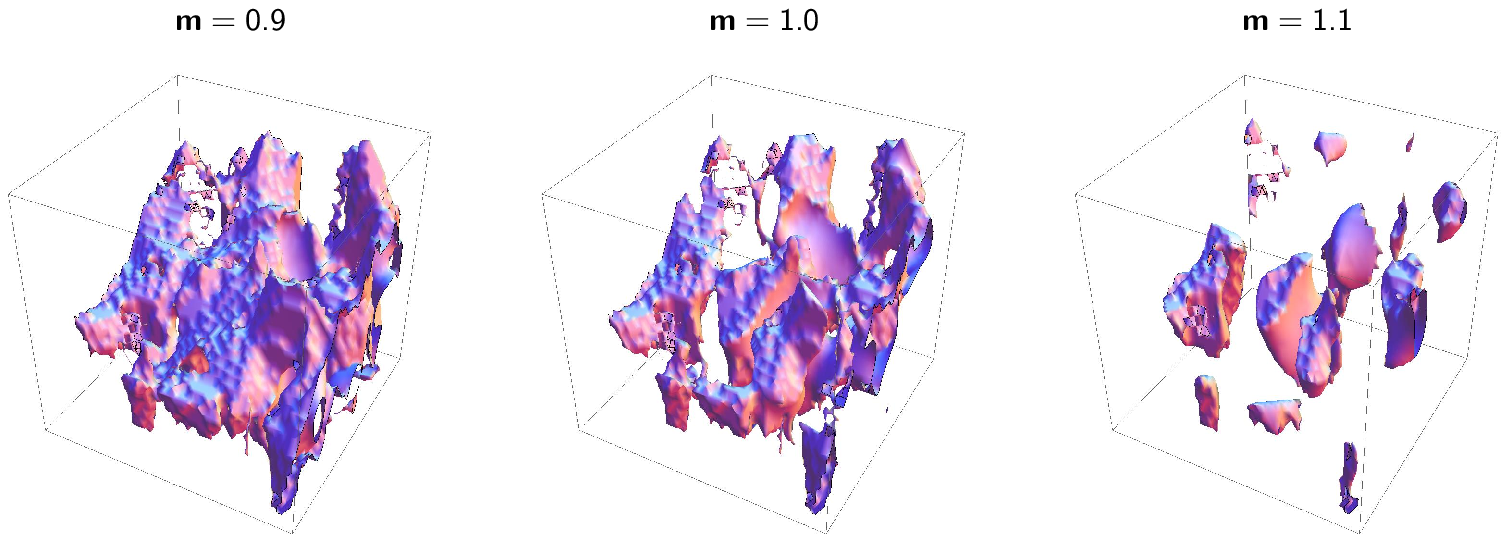}
\caption{Three contour plots, i.e. surfaces of constant model value,
 of the salt dome model  ($\m=0.9,1.0,1.1$).} \label{saltdome3dpic}
\end{figure}

\begin{figure}
\centering\resizebox{!}{20cm}{\includegraphics{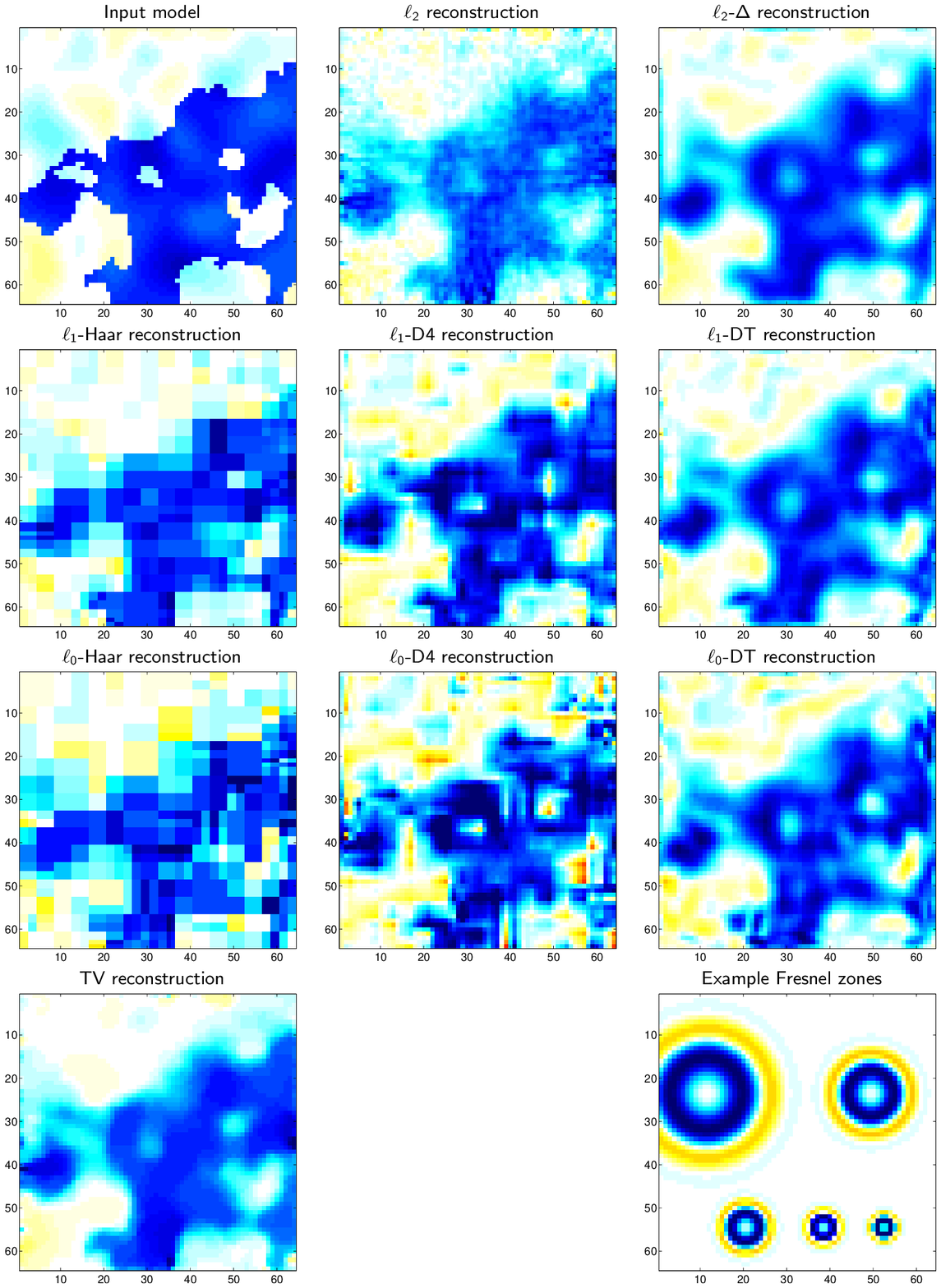}}
\caption{Horizontal slice number $25$ of the salt dome
model and its various reconstructions using all $24000$
kernels. The TV, $\ell_2$-$\Delta$, $\ell_1$-DT and $\ell_0$-DT methods
introduce the most smoothing. The $\ell_1$ and $\ell_0$ Haar reconstructions
have the least smoothing but are very blocky. A good compromise, in this
respect, may be found in the $\ell_1$-D4 reconstruction. The $\ell_0$-D4 reconstruction is less appealing. The bottom right figure shows
five representative kernel cross sections for different frequencies.
These cross sections are taken perpendicular to and in the middle of the central rays ,
and give the reader an idea of the size of the kernels relative to the structure in the model and
of the relative Fresnel zone widths of the different kernels.}\label{saltdomepic1blowup}
\end{figure}

\begin{figure}
\centering\centering\resizebox{!}{20cm}{\includegraphics{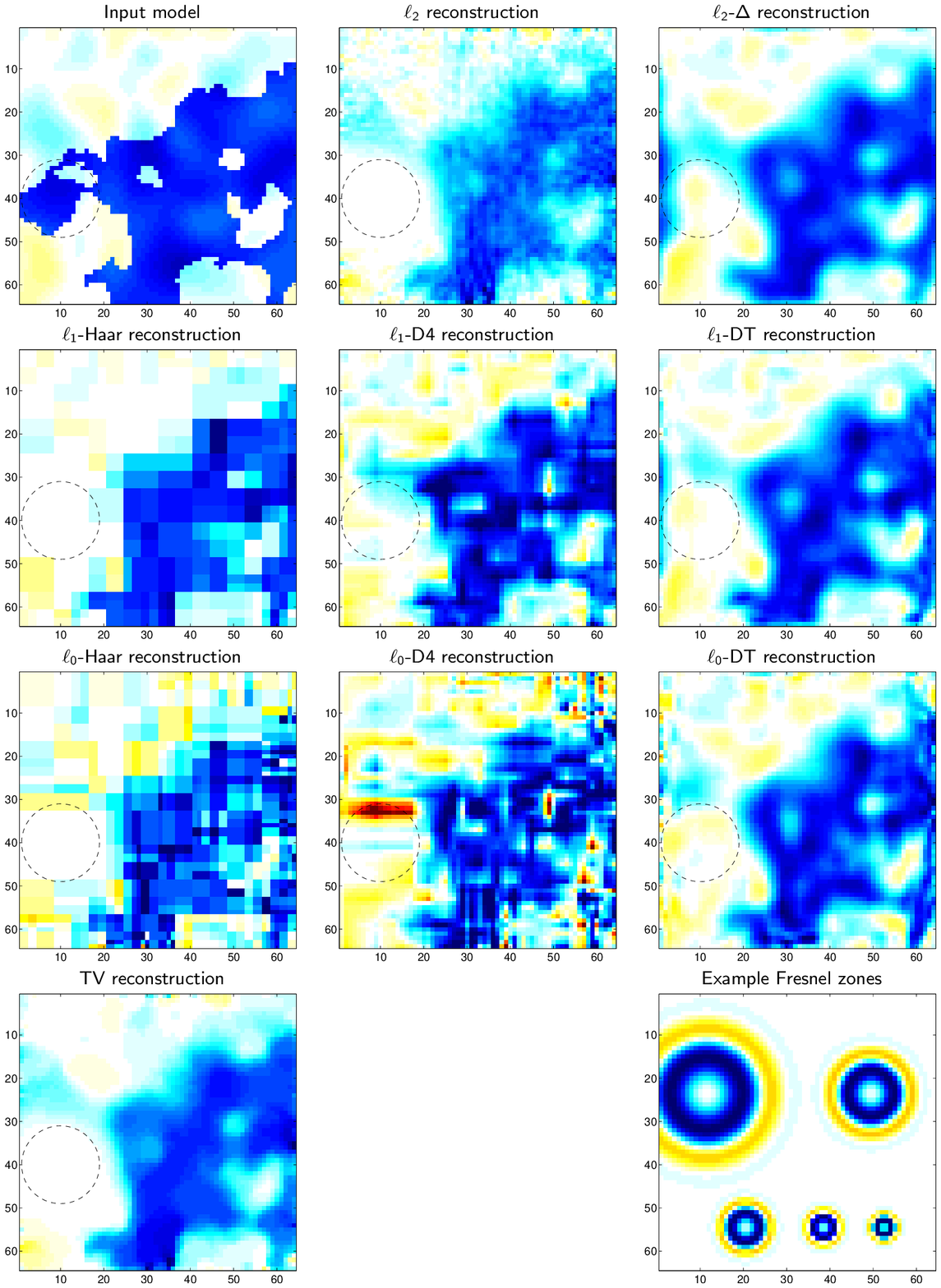}}
\caption{Horizontal slice number $25$ of the salt dome model and its reconstructions using only
$20000$ kernels and thus non-uniform coverage. The area of the input model that lies in the
region that is not covered by any kernel in this slice (indicated by a dashed circle), is not well reconstructed. Different
methods compensate for the missing information in different ways.
The $\ell_0$-D4 reconstruction shows a distinctive artifact. As in Fig.~\ref{saltdomepic1blowup},
the bottom right pane shows five representative kernel cross sections.}\label{saltdomepic2blowup}
\end{figure}

In this section we try to reconstruct a complex 3D model of a
realistic salt body in the subsurface. The complex salt dome
model was obtained from a prestack depth-migration of field
seismic data in the deep-water part of the Gulf of Mexico, and
was kindly provided to us by BP America, Inc. To better
accommodate the straight-ray tomography used in this paper, the
sediment velocities surrounding the salt dome model that were
present in the original model provided to us, were replaced
with a constant velocity. This model was superimposed on a
background model with long-wavelength variations (smoothed
Gaussian). The model has a rather sharp contrast between the
velocity in the salt and in the surrounding background model,
providing for sharp edges. A single horizontal slice through
the resulting salt dome model is pictured in Figures
\ref{saltdomepic1blowup} and \ref{saltdomepic2blowup}. For
completeness, all 64 horizontal slices are shown in Figure A.3
(right) of the online supplementary material (as well as the
smoothed Gaussian that was added in; Figure A.3, center). Three
contour plots of this model are shown in Figure
\ref{saltdome3dpic}, corresponding to the model values
$\m=0.9$, $\m=1$ and $\m=1.1$. The model contains $64\times
64\times 64=262144$ voxels as in the checkerboard examples.

We perform the same type of experiment as before: we construct
synthetic data and add $10\%$ Gaussian noise to it. From this
noisy data, we try to reconstruct the input model. There are
two differences with the checkerboard reconstructions. Firstly,
we will compare several different wavelet families for the
$\ell_1$ and $\ell_0$ reconstructions (in this case, there is
no obvious reason to prefer Haar wavelets over other wavelet
bases). Secondly, we will repeat the reconstruction experiment
for an operator that has only $20000$ kernels instead of
$24000$, to simulate imperfect coverage of the model domain by
the kernels. In other words, with the $20000$ kernel
reconstruction, a region of the model is ill resolved.

The wavelet families used are, in order of increasing
smoothness: Haar \cite{Haar1910}, D4
\cite{Cohen.Daubechies.ea1992,Cohen.Daubechies.ea1993} and
so-called directional dual tree (DT) wavelets
\cite{Kings1999,Kingsbury2002}. The Haar and D4 wavelet
transform on the cube are direct products of the corresponding
wavelet transforms in 1D. The DT wavelet transform is not and
it has, by construction, better directional sensitivity. The D4
wavelets that we will use do not suffer from edge effects as
they do not use periodic boundary conditions, but follow the
interval scheme proposed in
\cite{Cohen.Daubechies.ea1992,Cohen.Daubechies.ea1993}. Other
model parameterizations that could be used are shearlets or
curvelets (they are particularly suited to sparsely represent
models with singularities along curves or surfaces), but we did
not include them in our study
\cite{Labate.Lim.ea2005,Kutyniok.Labate2007,Candes.Demanet.ea2006}.

Judging the success of an algorithm to reconstitute the input
model invariably involves a degree of subjectiveness, even if
one designs a numerical measure for goodness of model fit. Such
measure might also depend on the goal of the scientific
experiment conducted. For example, if one deducts temperatures
from velocity variations, it is more important that the
amplitudes are correct and less important that sharp edges of
an anomaly are preserved, but a structural geologist may be
more interested in the edges and may wish to involve the misfit
of the gradient, for example.

In Table \ref{tablemisfit}, we list the amplitude misfit
($\|\m-\m^\mathrm{input}\|/\|\m^\mathrm{input}\|$) and judge
the fit to other features visually. A single horizontal slice
(number $25$) of the different reconstructions is pictured in
Figure \ref{saltdomepic1blowup} and all $64$ horizontal slices
are shown in Figure A.4 in the online supplementary material.
For the reconstructions using all $24000$ kernels, the TV
method works best based on the final resulting error, as well
as visual inspection. It is closely followed by the $\ell_1$
method using dual tree wavelets ($\ell_1$-DT) and by the
$\ell_2$ method with Laplacian smoothing ($\ell_2$-$\Delta$).
The $\ell_1$ method with D4 wavelets does better than with Haar
wavelets, that has a relative reconstruction error almost as
bad as obtained using the simple $\ell_2$ penalized method.
However, the ``top three'' methods (TV, $\ell_1$-DT,
$\ell_2$-$\Delta$) produce much smoother models than the input
model. In case the correct sharpness of features is a desirable
characteristic of the solution, the $\ell_1$ reconstructions
with Haar or D4 wavelets are more faithful to the input data.
In this case one may well prefer D4 over Haar to avoid the
rather blocky nature of the shapes. The qualitative differences
with the noisy $\ell_2$ reconstruction are obvious. The
$\ell_0$ reconstructions that were obtained using iterative
hard-thresholding, are less appealing. Numerically the
$\ell_0$-Haar and $\ell_0$-D4 perform worst of all
reconstructions, whereas $\ell_0$-DT comes in fifth.

The reconstructions with only $20000$ kernels are shown in
Figure \ref{saltdomepic2blowup} (a single horizontal slice) and
Figure A.5 in the supplementary material (all horizontal
slices). The most interesting comparisons are again done
visually. The lack of data coverage affects most strongly the
areas around voxel $(10,40)$ in slice 25 (Figure
\ref{saltdomepic2blowup}) and the lower left corner of slices
13--44 (third and fourth row in Fig. A.5 especially). Not
surprisingly, none of the algorithms is able to `recreate' the
model where there are no data at all. But close inspection of
the model near the edge of the region affected by the data gap
shows that the Haar and D4 wavelets produce the model that is
least contaminated by smoothing effects beyond the gap, with D4
occasionally trying to correctly `fill in'. The $\ell_0$-D4
reconstruction creates a distinctive artifact in this area. We
speculate that this is caused by the non-convex nature of the
$\ell_0$ problem (\ref{l0functional}).

\subsection{Computational aspects}

\label{compaspectssection}

Apart from the aspect of the visual reconstruction quality it
is also important to compare reconstruction times. The four
numerical algorithms that were used
---conjugate gradient, fast iterative soft and hard thresholding, and
reweighed conjugate gradient--- all need one application of
$\A$ and one application of $\A^T$ per iteration step. These
dominate the other, but much faster, operations such as
addition of vectors, vector norms etc that are also present in
each iteration step. The forward and inverse wavelet transforms
that are used in some of the methods (via the Fast Wavelet
Transform algorithm) also take a negligible time compared to an
application of $\A$ and $\A^T$. It follows that it is
sufficient to compare the number of iterations when we evaluate
the efficiency of different reconstructions.

The number of iterations and corresponding relative
reconstruction errors
$\|\m-\m^\mathrm{input}\|/\|\m^\mathrm{input}\|$ are given in
Table \ref{tablemisfit}. In all cases, the iterative
reconstruction algorithms were started from $\w^{(0)}=\zero$ or
$\m^{(0)}=\zero$.

For the checkerboard model the data in Table \ref{tablemisfit}
show that the $\ell_1$ and $\ell_0$ methods do extraordinarily
well, with a mean square error far below what could be expected
based on the data noise level of $10\%$. In the same sense, the
total variation minimization and the Laplacian penalization
perform somewhat better than the simple $\ell_2$ method; a
large number of simulations with different noise realizations
would be necessary to verify whether this is a statistically
significant difference. In case of the checkerboard
reconstructions, only $100$ iterations were performed for each
method. This shows that the $\ell_1$ and $\ell_0$ methods can
be very successful if the sought after model is very sparse in
the basis used, even with
a limited number of iterations.\\
We have also verified that the functionals
(\ref{l2functional}), (\ref{l2deltafunctional}),
(\ref{l1functional}), (\ref{l0functional}) and
(\ref{TVfunctional}) remain almost constant after this number
of iterations, as did the relative distance to the input model.

In case of the salt dome reconstructions, the $\ell_0$-Haar,
$\ell_0$-D4 and the simple $\ell_2$ method do worst (in terms
of reconstruction error) closely followed by the $\ell_1$-Haar
method. The latter is due to the inappropriateness of the Haar
basis to represent the salt dome model in a sparse fashion. The
other three methods ($\ell_1$-D4, $\ell_1$-DT and TV) do about
equally as the $\ell_2-\Delta$ method for the salt dome
reconstruction with $24000$ data.

To gauge wether there is a significant difference in
reconstruction error between the TV, $\ell_1$-DT and
$\ell_1-\Delta$ methods (and possibly the $\ell_0$-DT
technique), one would also need to repeat the numerical
experiment with many noise realizations.

The nonlinear reconstructions with $24000$ data were done with
$1000$ iterations. We used formula (\ref{fistabound}) to derive
a rough upper bound on the relative error remaining in the
functional (\ref{l1functional}) w.r.t to the minimum after this
number of iterations:
\begin{equation}
\frac{F(\w^{(n)})-F(\bar\w)}{F(\w^{(n)})}\leq
\frac{4\|\w^{(0)}-\bar \w\|^2}{\alpha (n+1)^2 F(\w^{(n)})}\lessapprox
10^{-3},
\end{equation}
where we used $\w^{(0)}=\zero$ and approximated $\|\bar\w\|$ by
$\|\w^{(n)}\|$. In other words, the calculated value of the
minimum of the functional is accurate up to three decimal
places (this bound is valid for the three wavelet families). A
simple plot of $F(\w^{n)})$ as a function of $n$ also reveals
that the functional is virtually constant after $1000$
iterations, a conclusion which also holds for the TV method,
the $\ell_0$ method and for the $\ell_2$ methods (after $100$
iterations).

The corresponding reconstructions with $20000$ data were done
with only $100$ iterations. In this way we demonstrate that a
reasonable result can already be obtained without an
excessively long computation time. This is evident by comparing
Figures \ref{saltdomepic1blowup} and \ref{saltdomepic2blowup}
(or Figures A.4 and A.5 in the supplementary material): apart
from the unresolved region near $(0.24,-0.7,-0.23)$ the
reconstructions are pairwise almost identical, despite the
significant differences in number of iterations. In other
words, the $\ell_1$ and TV algorithms already succeed, after a
small number of iterations, in producing qualitatively quite
characteristic reconstructions.\\
In case of the $\ell_0$ reconstructions the differences (far
from the unresolved region) are somewhat larger, we believe,
because the $\ell_0$ method only finds a local minimum of
(\ref{l0functional}).

As a result of the thresholding, the $\ell_0$ and $\ell_1$
algorithms provides sparse models at every iteration step (not
just in the limit $n\rightarrow\infty$). In other words it is
not necessary, or desirable, to run the FISTA/Nesterov style
algorithm for a very long time. Even after a small number of
iteration, they will provide a sparse model that fits the data
to within its error bars.

\begin{table}
\centering
\begin{tabular}{l|rr|rr|rr}
& \multicolumn{2}{c|}{Checkerboard} &\multicolumn{2}{c|}{Saltdome w. $24000$ data} &\multicolumn{2}{c}{Saltdome w. $20000$ data} \\
                  & Iterations & Error (\%) & Iterations & Error (\%) & Iterations & Error (\%) \\ \hline
$\ell_2$          & 100  & 68.8 & 100   & 48.8  & 100  & 56.9\\
$\ell_2$-$\Delta$ & 100  & 61.6 & 100   & 40.1  & 100  & 46.4\\[1mm]
$\ell_1$-Haar     & 100  & 1.8  & 1000  & 48.1  & 100  & 55.1\\
$\ell_1$-D4       & --   & --   & 1000  & 42.8  & 100  & 50.0 \\
$\ell_1$-DT       & --   & --   & 1000  & 39.5  & 100  & 46.8\\[1mm]
$\ell_0$-Haar     & 100  & 4.4  & 1000  & 61.5  & 100  & 67.9\\
$\ell_0$-D4       & --   & --   & 1000  & 52.6  & 100  & 63.2\\
$\ell_0$-DT       & --   & --   & 1000  & 44.1  & 100  & 49.4\\[1mm]
TV                & 100  & 64.0 & 1000  & 39.0  & 100  & 49.7\\
\end{tabular}
\caption{The number of iterations and the resulting
reconstruction error for the various models and
methods.}\label{tablemisfit}
\end{table}

\section{Conclusions}

In Paper I we showed how a large scale anomaly could be
reconstituted even where it was ill resolved because of the
selective nature of the wavelet coefficients and the $\ell_1$
criterion: one wavelet coefficient reconstituting a large,
circular, anomaly gave a better optimization than a couple of
coefficients reconstituting only the resolved part. With the
results of the much more complex salt dome model at hand, we
must now conclude that this probably represents more a (lucky)
exception than a rule. There is no magical solution for the
absence of data.

For the checkerboard reconstructions, the $\ell_1$ method with
Haar wavelets is able to do very well ---much better than could
be expected based on the data themselves--- because the Haar
wavelets are very efficient in representing \emph{this
particular} checkerboard pattern in a sparse way. The success
of the $\ell_1$ method thus depends heavily on the choice of a
suitable basis. For realistic models it is much more difficult
to find a good ---sparsifying--- basis, and the reconstruction
errors will be much larger. For the 3D salt dome
reconstruction, one could argue that the $\ell_1$-DT method
does well because it has good directional sensitivity and is
therefore able to adapt to the ``curvy'' nature of the outline
of the salt body, as opposed to $\ell_1$-Haar and $\ell_1$-D4
methods. The $\ell_2$-$\Delta$ method does well because the
Gaussian background that is present in the model is smooth
`noise' and this is exactly the prior information put into the
minimization criterion. The TV method does well as the main
part of the salt dome model is roughly piecewise constant and
TV favors that. The $\ell_0$ methods do not perform
particularly well, both from a quantitative as a qualitative
side.

The wavelets, however, do have the distinctive quality of
retaining sharp features even when regularizing by penalizing
highly oscillatory models. If the preservation of sharp
boundaries is not as important as the correct estimation of
amplitudes, the smoothed solution, using the $\ell_2$-$\Delta$
method, is to be preferred as it is fully linear and efficient
to solve with conjugate gradients. Methods using wavelets with
small support, however, are able to retain sharp features,
despite their regularization effect that penalizes highly
oscillatory models. These methods are thus preferable when
edges are important; our preference would go to the $\ell_1$-D4
algorithm which gives less blocky solutions than $\ell_1$-Haar.
In no cases should one use simple norm damping ($\ell_2$
method). Without imposing additional smoothing, i.e. while
still allowing for sharp transitions, the $\ell_1$ methods
yield models which do not show signs of noise.

The $\ell_0$ methods, which use hard-thresholding of wavelet
coefficients rather than soft-thresholding, cannot outperform
the $\ell_1$ methods. In some cases they appear to produce
severe artifacts. Another reason not to favor $\ell_0$
penalties is that they only produce a local minimum of the
functional (\ref{l0functional}). This may lead to larger
variability in the reconstructions (depending on the starting
point of the iteration). There is currently no proven technique
to tackle the minimization more efficiently that algorithm
(\ref{iht}). The (unproven) method (\ref{fiht}) proposed in
this paper is, as far as the authors can tell, new.

We conclude that using hard-thresholding is less appealing than
using soft-thresholding of wavelet coefficients: the
mathematical theory is less developed, the hard-thresholded
reconstruction may exhibit significant artifacts and the
reconstructions are not better than the ones obtained with
soft-thresholding ($\ell_1$ method).

Speed-wise the nonlinear methods cannot do better than the
conjugate gradient algorithm for the $\ell_2$ methods. Many
applied mathematics groups
\cite{Figueiredo.Nowak.ea2008,Bioucas-Dias.Figueiredo2007,%
Elad.Matalon.ea2007a,Kim.Koh.ea2007,Beck.Teboulle2008,%
Hale.Yin.ea2008,Yin.Osher.ea2008} are currently working on
speeding up the iterative soft-thresholding algorithm
(\ref{ist}), but it is still at least as time-consuming to use
the $\ell_1$ norm as it is to use the $\ell_2$ norm for
penalization, especially for severely ill-conditioned matrices
and low noise conditions \cite{Loris2009}.

In case the data is heavily contaminated by noise, it follows
from relation (\ref{penaltychoice}) that a large value of the
penalty parameter $\mu$ must be chosen. In \cite{Loris2009} it
was demonstrated that many competing algorithms for minimizing
an $\ell_1$ penalized functional converge quickly in such a
case. We therefore expect that such methods remain competitive
with the traditional $\ell_2$ smoothing methods in case of
travel time seismic tomography where the data noise level may
reach 50\%.

Based on the results in this paper, we can conclude that the
nonlinear methods offer a way to invert data and denoise the
resulting model in a single procedure without necessarily
smoothing the model too much. The two salt dome examples also
show that a good reconstruction, clearly showing the
characteristic effects of the penalizations used, is still
possible with a very limited number of iterations: This is a
consequence of the FISTA algorithm producing sparse models at
every iteration step. Sparse models can therefore be
constructed with few iterations and little computer time (see
\cite{M.C.1987} for a discussion of the number of iterations
used as a regularization parameter).

As an alternative to D4 or complex DT wavelets one could
consider using curvelets or shearlets, as they are naturally
designed to sparsely represent singularities along smooth
curves, such as, e.g., the sediment salt interface in our
model. In this work we have not studied how the different
regularization methods behave in conjunction with these
particular choices of dictionaries.

\section{Acknowledgments}

The authors would like to thank BP America Inc. for kindly
providing the 3D salt dome velocity model and several referees
for their suggestions that helped improve the manuscript. Part
of this research has been supported by the Francqui Foundation
(IL), the VUB-GOA 062 grant (ID, IL), the FWO-Vlaanderen grant
G.0564.09N (ID, IL) and NSF grant DMS-0530865 (ID).


\end{document}